# FEL stochastic spectroscopy revealing silicon bond softening dynamics

Dario De Angelis[1]* Emiliano Principi[1], Filippo Bencivenga[1], Daniele Fausti[1,2,3], Laura Foglia[1], Yishay Klein[4], Michele Manfredda[1], Riccardo Mincigrucci[1], Angela Montanaro[1,2], Emanuele Pedersoli[1], Jacopo Stefano Pelli Cresi[1], Giovanni Perosa[1], Kevin C. Prince[1], Elia Razzoli[5], Sharon Shwartz[4], Alberto Simoncig[1], Simone Spampinati[1], Cristian Svetina[5], Jakub Szlachetko[6], Alok Tripathi[4], Ivan A. Vartanyants[7], Marco Zangrando[1,8], and Flavio Capotondi[1]

[1]*Elettra-Sincrotrone Trieste, Strada Statale 14-km 163.5, Basovizza, 34149, Trieste, Italy*
[2]*Department of Physics, Università degli Studi di Trieste, 34127, Trieste, Italy*
[3]*Lehrstuhl für Festkörperphysik, Friedrich-Alexander-Universität Erlangen-Nürnberg, Erlangen 91058, Germany;*
[4]*Physics Department and Institute of Nanotechnology and advanced Materials, Bar Ilan University, Ramat Gan, 52900, Israel*
[5]*Paul Scherrer Institut, CH-5232 Villigen, Switzerland*
[6]*SOLARIS National Synchrotron Radiation Centre, Jagiellonian University, Czerwone Maki 98, 30-392 Krakow, Poland*
[7]*Deutsches Elektronen-Synchrotron DESY, Notkestr. 85, 22607 Hamburg, Germany*
[8]*Istituto Officina dei Materiali, CNR, Strada Statale 14-km 163.5, Basovizza, 34149, Trieste, Italy*

Time-resolved X-ray Emission/Absorption Spectroscopy (Tr-XES/XAS) is an informative experimental tool sensitive to electronic dynamics in materials, widely exploited in diverse research fields. Typically, Tr-XES/XAS requires X-ray pulses with both a narrow bandwidth and sub-picosecond pulse duration, a combination that in principle finds its optimum with Fourier transform-limited pulses. In this work, we explore an alternative experimental approach, capable of simultaneously retrieving information about unoccupied (XAS) and occupied (XES) states from the stochastic fluctuations of broadband extreme ultraviolet pulses of a free-electron laser. We used this method, in combination with singular value decomposition and Tikhonov regularization procedures, to determine the XAS/XES response from a crystalline silicon sample at the $L_{2,3}$-edge, with an energy resolution of a few tens of meV. Finally, we combined this spectroscopic method with a pump-probe approach to measure structural and electronic dynamics of a silicon membrane. Tr-XAS/XES data obtained after photoexcitation with an optical laser pulse at 390 nm allowed us to observe perturbations of the band structure, which are compatible with the formation of the predicted precursor state of a non-thermal solid-liquid phase transition associated with a bond softening phenomenon.

* deangelis@iom.cnr.it

# I. INTRODUCTION

Over the last decades, the high brightness achievable by X-ray synchrotron radiation light sources, together with their high energy resolution and control of the polarization, gave an impressive boost to techniques capable of retrieving electronic and structural properties of matter. Nowadays, synchrotron spectroscopy is extensively used in a large number of scientific fields, ranging from condensed matter, organic and inorganic chemistry, nanotechnology, photo-catalysis and molecular biology. As an example, it is now routine to trace chemical reactions with XAS and XES approaches, map the spin states via resonant effects in the dichroic response (e.g. Faraday or Kerr rotation), or detect molecular chirality [1–3].

More recently, time-resolved approaches combining synchrotrons and optical lasers have enabled the investigations of dynamic processes on a time-scale of tens of ps or longer [4–7], and generally with very low energy synchrotron pulses. With the advent of ultrafast (<100 fs) sources like High Harmonic Generation (HHG) laboratory lasers and Free Electron Lasers (FELs), shorter time scales and higher pulse energies have become available. This has facilitated access to the dynamics of electronic band excitations [8–11], and high-resolution (tens of meV) Tr-XAS/XES represents a strategic asset in the investigation of fundamental science, as well as for characterization of functional materials and catalytic applications [12, 13].

HHG based techniques suffer from low intensity at higher photon energy, such as the soft x-ray region. FELs provide much higher photon fluxes, but there are few sources, so that access is limited. The efficient use of FEL sources requires the implementation of novel approaches to spectroscopy, which benefit from the high peak brightness and broad spectral bandwidth typical of FEL pulses. In this respect, it is highly desirable to develop multiplexing approaches capable of simultaneously exploiting a large spectral bandwidth without the need for spectral filtering and photon energy scanning. An elegant and effective solution to multiplex-based spectroscopy is represented by correlation spectroscopy methods based on the determination of the statistical fluctuations introduced by the interaction of the light pulses with the sample; such approaches have been realized both in the optical region [14, 15] as well as in the X-ray range [16–19].

Here we extend those techniques to the EUV regime, and in particular to the $L_{2,3}$-edge (photon energy ≈100 eV), to investigate the structural and electronic dynamics of crystalline silicon (c-Si).

c-Si exhibits different photo-induced dynamics depending on the optical pump intensity and wavelength, especially if the excitation is above or below the direct conduction-valence band energy gap. At low fluence (<10 mJ/cm$^2$) [20–22], i.e. when the laser deposited energy does not modify the covalent nature of c-Si bonding and no phase transition is excited, the early response of the electronic structure to an external optical stimulus is characterized by the modification of the band structure profile and density of states. State-blocking based on the Pauli principle, broadening, and band gap renormalization narrowing effects have been identified on the sub-ps time scale [20]. At longer times, of the order of tens of ps, i.e. when electron-phonon coupling transfers the electronic excitation energy to the crystal lattice, structural deformations, such as directional and isotropic thermal expansion, have been reported as additional relaxation processes [21, 22]. At high fluence (> 170 mJ/cm$^2$), when the deposited energy per atom is sufficient to modify the covalent nature of c-Si bonding, various experimental works indicated that the optical stimulus drives the c-Si band structure across a solid-liquid phase transition, occurring on the few ps time scale [23–26]. Extensive investigation of non-thermal melting, from both theoretical [27–31] and experimental standpoints [32–34] have shown that after a few tens of ps, the phase transition proceeds through a polymorphic evolution of the liquid phase, from an initial high-density disordered structure to another liquid phase with lower atomic coordination number [29, 35]. At moderate fluence, i.e. before the onset of the solid-liquid phase transition, it has been predicted that a precursor state of the non-thermal melting process can be identified in the softening of interatomic bonds. This is described as a modification of the potential energy landscape, that induces the instability of transverse and longitudinal acoustic phonons [27, 36, 37]. This Bond Softening (BSo) phenomenon has been experimentally observed for several classes of bulk materials after excitation with sub-ps optical pulses, as a fluence-dependent temporal modulation of the intensity of Bragg spots in x-ray diffraction [38–41]. On freestanding polycrystalline Si membranes, an exponential decay of the X-ray diffraction signal was first observed and interpreted as a heat-transfer dynamic on the atomic scale, characterized by a well-defined time constant [42]. This heating dynamics was later described by an ab-initio simulation [36] in terms of a BSo. These authors calculated that the modification of the potential energy landscape proceeds in a c-Si system through thermal phonon squeezing when excited below the Lindemann stability limit [43]. Moreover, they noted that the time constant associated with the phonon squeezing process depends on the number of electron-hole pairs per unit volume that are generated by the laser pulse.

In this work, the combined XAS/XES analysis allowed us to determine the BSo on the timescale of electronic thermalization.

## II. RESULTS

### A. Experimental setup

The XAS/XES measurements were performed at the EIS-TIMEX beamline of the FERMI FEL facility in Trieste (Italy) [44]. The beamline is optimized for pump-probe measurements; two spectrometers are available along the photon beam path (one upstream and the other downstream of the experimental chamber) and both of them can operate in single-shot mode at the FEL repetition rate (50 Hz). The first spectrometer, called PRESTO [45], is integrated in the FERMI photon transport system. It allows us to characterize the spectral content of each FEL pulse by imaging, in the energy dispersive plane, the first diffraction order of a variable line spacing grating. The second one, called WEST [46], is placed downstream of the TIMEX end-station and is designed to collect the FEL photons transmitted by the sample. It comprises an EUV reflective grating and detector. The resolving power is $1 \cdot 10^4$ for PRESTO and $4 \cdot 10^3$ for WEST. The sample was an ultra-polished c-Si (001) p-doped membrane, 200 nm thick, purchased from Norcada.

We used the FERMI FEL-2 source [47] that was operated in the optical klystron (so-called "SASE-like") mode [48]. This mode maximizes the radiation bandwidth, while achieving the stochastic ("spiky") spectral profile required by our analysis method (see supplementary information). To fully cover the c-Si $L_{2,3}$ absorption edge, we scanned the photon energy of the SASE beam in 5 steps from 99.0 eV to 102.5 eV. Each step covered a distinct sub-region of the absorption edge, with average central photon energies of 99.6, 99.8, 100.5, 101.0, and 101.6 eV. The SASE-like emission bandwidth was approximately 0.8 eV.

### B. Static sample characterization

As a preliminary step, we performed XAS/XES measurements across the $L_{2,3}$-edge of c-Si. The results are shown in Figure 1, where the colored boxes correspond to the different sub-regions of photon energy. As described in detail in the Supplemental Material, to retrieve information on the elastic (XAS) and the inelastic (XES) X-ray response of the sample, we collected $10^5$ spectra for each of the aforementioned sub-regions of the photon energy range. The sample response was evaluated using the Tikhonov regularization method [49] involving the transfer matrix connecting the ensemble of input and output spectra. These results were obtained with five acquisitions of about 20 min each at 50 Hz FEL repetition rate, and the time needed to change the FEL photon energy range was about 2 minutes. The data analysis can be carried out on-line right after data acquisition.

In order to maximize the signal to noise ratio in the PRESTO and WEST spectrometers, we discard the FEL shots with very low intensity, i.e. when the recorded intensity on a spectrometer was barely visible above the noise level. In post processing filtering analysis, we calculated the average and the root mean square (rms) spectral intensity of the full dataset and applied a threshold filter to disregard all shots whose intensity was more than two times the variance lower than the average, i.e. approximately 2% of the total number of shots.

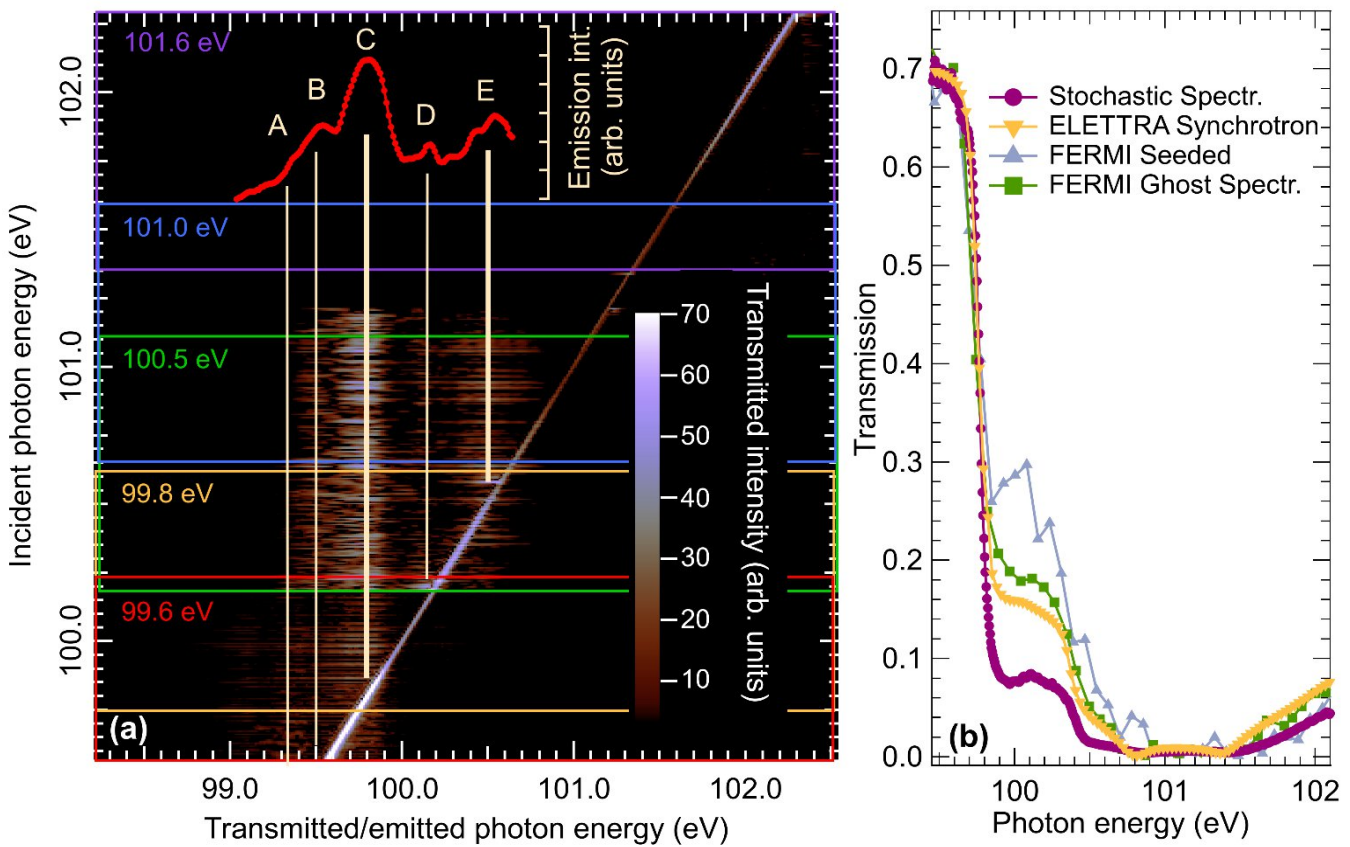

FIG. 1. (a) Reconstructed sample response map at equilibrium. Colored rectangles refer to the 5 photon energy settings of the SASE-like emission (the text labels with the same color indicate the average photon energy for each setting), while vertical white lines indicate the emission features. The inset in red represents the integrated emission spectrum; letters from A to E label the emission features. (b) Transmission profile of c-Si, as obtained from the diagonal terms of the sample response map (purple dots connected by lines, here error bars are < 2% and cannot be resolved on this scale), is compared to other spectroscopic methods, as detailed in the legend.

In Figure 1(a), clear intensity modulations along the matrix diagonal are distinguishable. This corresponds to the sample response at the energy of the input radiation, i.e. the XAS spectrum across the c-Si edge.

Additional off-diagonal vertical features are observable in the upper triangular matrix. Those features correspond to X-ray emission processes as a result of electronic transitions from higher, occupied states into the core-holes created by input EUV photons. Emission lines are visible at about (A) 99.35 eV, (B) 99.50 eV, (C) 99.76 eV, (D) 100.10 eV, and (E) 100.45 eV. Lines C and E can be associated with the radiative de-excitation into the $L_2$ and $L_3$ orbitals. The complex X-ray emission spectrum is governed by multi-electron interaction and core-hole screening effects, resulting from participation of the intermediate state in excitation and de-excitation processes. The faint series of emission lines at lower energies are in good agreement with the structures observed in Si nanoparticles by Šiller and co-workers [50]. They ascribe those features to Si in-gap states due to defects or doping, e.g. oxygen or carbon impurities, resulting in deeper donor states located ≈0.25 eV below the conduction band minimum.

Figure 1(b) reports the integrated profile along the matrix diagonal (purple dots connected by lines). The spectrum clearly displays two EUV transmission drops assigned to the spin-orbit-split resonances of the 2p levels at 99.76 eV ($L_3$-edge) and 100.45 eV ($L_2$-edge), corresponding, respectively, to emission features C and E. The present data are compared with:
i) a spectrum from a nominally identical sample obtained by a synchrotron measurement (at the BEAR beamline of the Elettra synchrotron [51], yellow line) with comparable energy resolution and acquisition time;
ii) a measurement performed at FERMI by using narrowband (seeded mode) FEL emission and scanning the photon energy (grey line). It is worth noticing the lower energy resolution of about 75 meV due to the reduced number of experimental points collected in the same acquisition time (2 hours);
iii) a measurement performed at FERMI (in SASE-like mode) obtained by exploiting the ghost spectroscopy method (green line) [18]: the energy and temporal resolution, as well as the acquisition time, are comparable.

Besides the remarkable data quality, we observed different ratios among the relative amplitude of the $L_2$ and $L_3$ edges in the four cases. This can be ascribed to the high sensitivity of this parameter to the chemical and structural environment of the Si atoms [52–55], added to the fact that the measurements were performed on different Si membranes, although nominally identical, that may have been affected by different contamination levels.

This comparison validates the correlation spectroscopy method of Kayser et al. [16] in the EUV regime, indicating that the energy resolution is superior to FEL data acquired by scanning the photon energy and comparable to standard synchrotron measurements operated with an X-ray monochromator. In addition, the present approach allows for the simultaneous characterization of occupied and unoccupied electronic states, thus providing more comprehensive information on the electronic configuration of the scattering system.

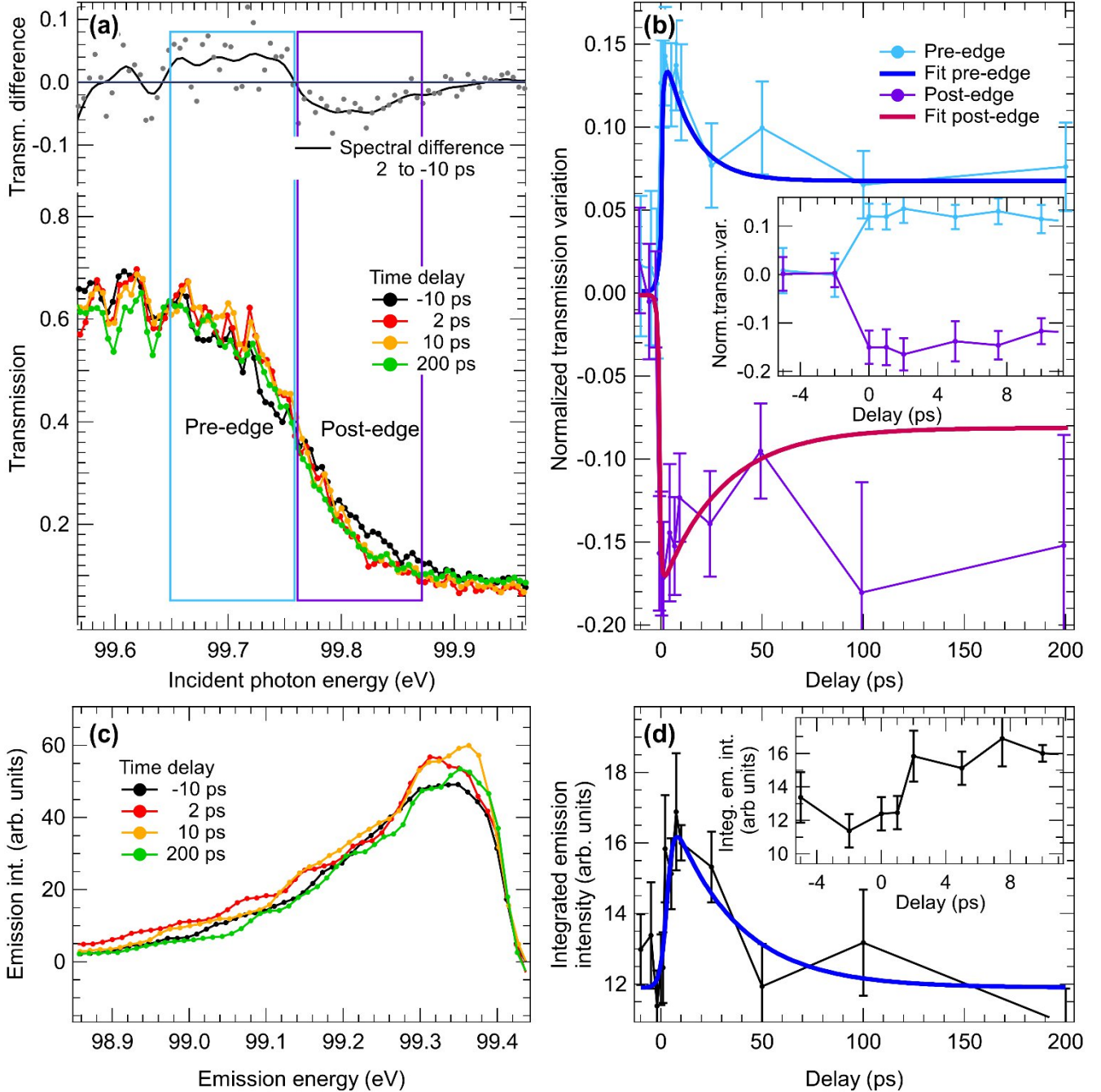

FIG. 2. (a) Selected transmission spectra before and after optical excitation, as indicated in the legend. The light blue and purple rectangles define the "pre-edge" (99.65 – 99.76 eV) and "post-edge" (99.76 – 99.87 eV) spectral regions. (b) Light blue and purple data refer to the time dependence of the normalized variation of the transmitted intensity in the pre-edge and post-edge regions, respectively; fit results are shown in the same plot; the inset shows an enlargement of the initial part of the dynamics. (c) Profiles of the emission line at 99.35 eV as a function of the delay, as indicated in the legend. (d) Time dependence of the total emission intensity at 99.35 eV: experimental data in black, fit result in blue. The inset shows an enlargement of the initial part of the dynamics.

Moreover, an important advantage with respect to synchrotron-based experiments, is the use of an ultrafast XUV source, that can be used for time-resolved sub-ps spectral investigations.

### C. Laser driven dynamics

We used EUV correlation spectroscopy in a time resolved experiment to investigate the dynamics of c-Si after fs optical excitation with 20 mJ/cm$^2$ fluence. This regime is below the damage threshold and has been theoretically described in the work of Zijlstra et al. [36] as a precursor to a non-thermal melting process. To photo-excite the c-Si sample, we employed an optical pulse with 3.16 eV photon energy (392 nm wavelength) and 90 fs pulse duration. A pump-probe time-delay scan consisted of about 20 time-delay points, from -5 to 250 ps, and at each point we acquired a dataset consisting of 50,000 SASE-like FEL shots. The FEL pulse duration was about 280±30 fs in full width at half maximum and it was the factor limiting the time resolution of our experiment. We focused on the Si $L_3$-edge, and we set the nominal SASE photon energy at 99.6 eV. The time needed to complete a scan was about eight hours, and at such moderate fluence levels, no sample damage during this time was observed during the experiment. To obtain Tr-XAS/XES profiles, the correlation analysis was performed for each time-delay dataset (more details in the Supplemental Material). Figure 2(a) shows the retrieved XAS profile at four selected time delays: before (-10 ps) and after (+2, + 25, + 200 ps) the optical excitation.

Two main features characterize the dynamics of the XAS spectra from the excited sample, namely an increase in the sample transmission before the absorption resonance and a decrease after it. Figure 2(b) reports the dynamics of the normalized XAS at the $L_3$-edge; the inset in Figure 2(b) shows an enlargement of the first 10 ps dynamics. We can identify two distinct dynamics: (i) a rise (decrease) of the sample transmission in the pre- (post-) edge area in the first 2 ps after the excitation; (ii) an inflection (increase) of the transmission between 2 ps and 50 ps, slightly faster in the pre-edge region with respect to the post-edge one, which progressively recovers the transmission of the sample before excitation.

To estimate the characteristic time constant for these dynamics, we fitted the two datasets with a model function defined as a logistic function attenuated by an exponential factor for positive delay values. From these fits, the exponential attenuation time is 15 ± 5 ps for the pre-edge and 20 ± 5 ps for the post-edge. The rise (fall) time is beyond the intrinsic time resolution of our measurement, determined by the FEL pulse duration.

A dynamic effect was concurrently observed in the XES line at 99.35 eV (feature A), as displayed in Figure 2(c). Here selected emission spectra (integrated over the vertical energy dimension in Figure 1(a)) are shown. The corresponding dynamics are reported in Figure 2(d), where we observed an increase in the emission intensity in the first 10 ps after the excitation and a slower recovery on longer time scales. To analyze this dataset, we used the same fitting function described above: the rise time appears longer than the one observed in the transmission variation (2 ± 0.5 ps), while the recovery time constant is 22 ± 5 ps, which is similar to the time constant we observed in Tr-XAS.

### III. DISCUSSION

The electronic and structural dynamics induced in Si by short laser pulses have been extensively investigated previously. Depending on the excitation fluence, different dynamical regimes have been recognized. At low fluence, below 10 mJ/cm$^2$, thermal phenomena are more prominent than electronic ones, as discussed by Cushing et al. [22], where EUV spectroscopy is used to describe the hot phonon and carrier relaxation in a c-Si excited by different femtosecond laser pulse in the range from infrared to ultraviolet. At 800 nm pumping wavelength, the authors observed a broadening of the absorption edge, which, to make a direct comparison with our results, can be described as a decrease (increase) of sample transmission in the pre-edge (after-edge) spectral region. They also modeled this effect using a density functional theory approach. The observed dynamics was rationalized in terms of state-filling, band gap renormalization and lattice deformation. However, in the present work we observe the opposite behavior of the $L_3$-edge (i.e. a narrowing of the edge after the excitation), indicating that the framework described by Cushing et al. is not compatible with our experiment, and that the larger pumping fluence in our case gives rise to non-thermal phenomena.

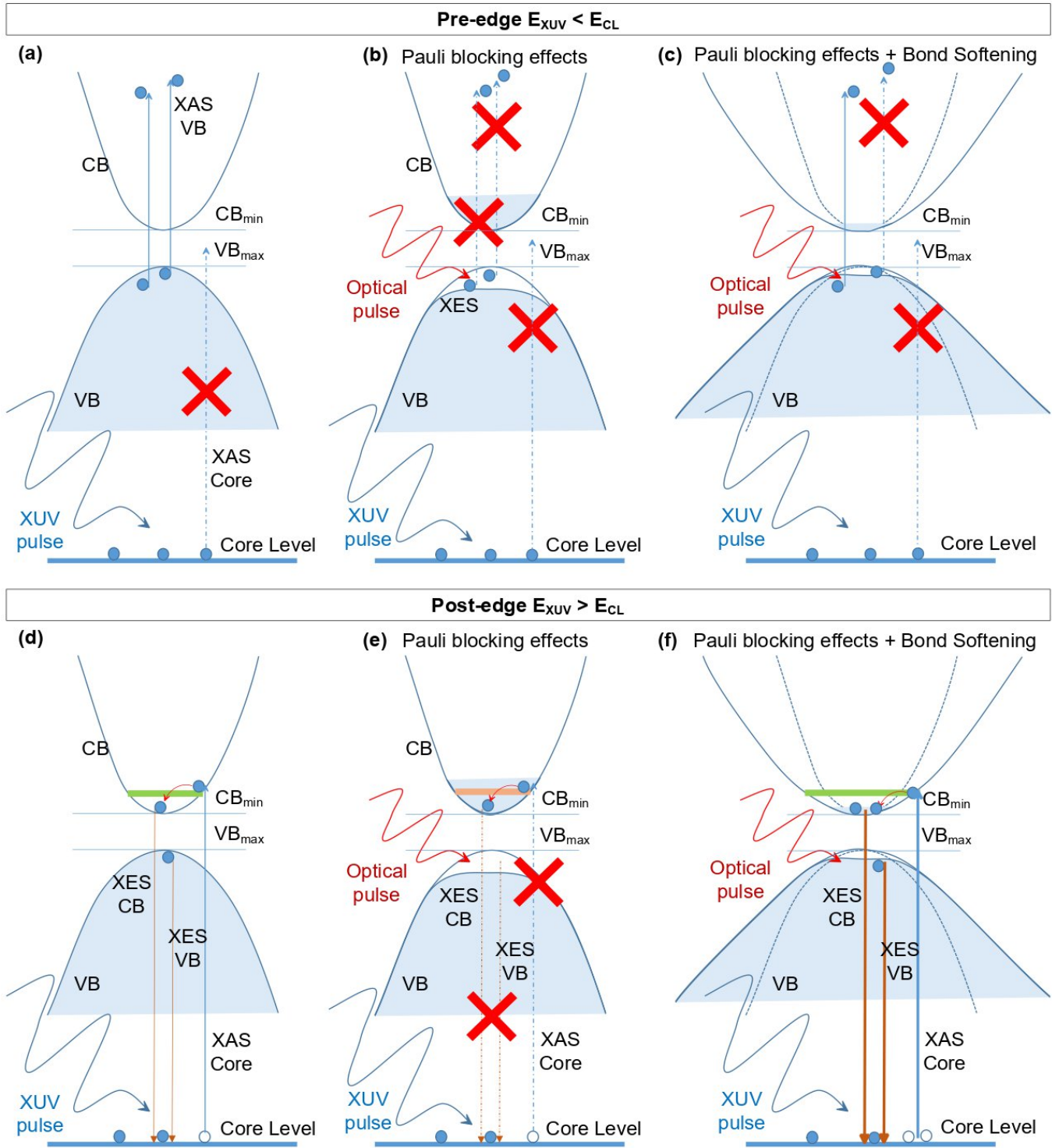

FIG. 3. Schematic representation of the electronic processes described in the text, related to XUV absorption. (a) and (d): static case, i.e. XUV absorption in the absence of the laser pump pulse, for XUV energies below and above the absorption edge (pre-edge and post-edge) respectively. (b) and (e) XUV absorption occurring after laser pump, in the absence of the BSo effect, for XUV energies below and above absorption edge respectively). (c) and (f) XUV absorption occurring after laser pump in the presence of the BSo effect, for XUV energies below and above absorption edge respectively.

On the other hand, melting phenomena start to take place for pump fluence above 200 mJ/cm$^2$. In this condition, Beye and co-workers [32] induced a non-thermal melting of Si in the first 2 ps, followed by a phase transition from a low-density liquid (LDL) to a high-density liquid (HDL) phase, observed 5 ps after the excitation. It is worth noting that the fluence used in that experiment is one order of magnitude higher than that used in the present experiment. At the fluence employed in the present experiment, we did not observe evidence of melting phenomena, neither from transmission measurements nor from microscopic inspection, despite the prolonged (several hours) exposure to the laser and FEL beams. This conclusion is endorsed by the experimental observation (by microscopic inspection and transmission measurement) of no visible sample damage even after eight hours of exposure to the pump pulse.

We exclude these two opposing pump fluence regimes as valid interpretative frameworks and consider the calculations and the experimental observations reported by Zijlstra et al. [36]. On the basis of the excitation fluence regime described there, we can infer that our pumping conditions were suited to induce BSo, i.e. the weakening of the potential energy surface that indirectly generates an increase in the electronic density of states close to the conduction band minimum. Those processes are schematically represented in Figure 3 for pre-edge ((a), (b), (c)) and post-edge ((d), (e), (f)) photon energy condition respectively, under the hypothesis that the band gap energy (i.e. the energetic difference between the conduction band (CB) minimum and the valence band (VB) maximum) is unaffected by optical pumping.

Panels (a) and (d) describe the static situation and the main processes considered. At the pre-edge energy (Fig. 3(a)), i.e. for photon energy ($E_{ph}$) smaller than the energy of the core level (CL) transition ($E_{CL}$), only electrons from the VB can be excited above the $CB_{min}$, contributing to the XAS signal. On the other hand, since $E_{ph}$ is smaller than $E_{CL}$, the absorption from the core level is suppressed. As a consequence, no XES signal can be expected, due to the lack of a core hole required for such an electronic transition. The static situation is more complex for the post-edge energy ($E_{ph}> E_{CL}$) (Fig. 3(d)). Here, additionally to the processes described for the pre-edge case (not shown in Fig. 3(d)), the higher photon energy activates a new absorption channel from the core level to the conduction band. The electrons excited by such an electronic transition, ultimately limited by the density of states in the CB, can quickly (<10 fs) de-excite to $CB_{min}$ by means of inelastic intra-band energy loss processes. On the same time scale, the electronic vacancy in the core level (core-hole), generated by the absorption process, can be filled by the spontaneous emission of secondary photons (XES) involving either CB or VB electrons.

The Pauli blocking effects [55, 56] induced by optical pumping (with photon energy larger than the band gap) are described in Figures 3(b) and 3(e) for pre-edge and post-edge situations, respectively. In these panels, no BSo is taken into account and no deformation of the VB and CB profiles is considered. Under such assumptions, for the pre-edge case (Fig 3(b)), the optical pump will excite electrons from the top of the VB to the bottom of the CB, and such a process is competitive with XUV absorption, indeed the laser pump will fill empty states required for the XUV electronic transition. As a consequence, an increase of XUV sample transmission is expected (graphically represented by crossed dash-dot lines in Fig. 3(b) with respect to the stationary condition, (Fig. 3(a)). The same effects are expected for the post-edge case depicted in Fig 3(e). In this case the laser pulse will populate states in the conduction band inducing a suppression of core level absorption and, as a consequence of the reduced number of final states, a reduction of the XES signal. While the Pauli blocking effect phenomenologically reproduces the experimental observable in the XAS signal for the pre-edge condition, the opposite trend is observed for photon energy above the Si $L_{2,3}$ edge in the experimental XAS and XES time resolved signals (Fig. 2(a) and 2(c)).

The effects of BSo in the electronic transition, after optical excitation, are represented in Fig. 3(c) and 3(f) for pre-edge and post-edge XUV photon energy, respectively. In these figures, the BSo is represented as a weakening of the CB and VB energetic gradient (continuous blue lines for BSo case, dashed lines for static case). Such a variation of band profiles increases the density of the available states both above the $CB_{min}$ and below the $VB_{max}$. As a consequence, for the pre-edge case (Fig 3(c)) the optically excited electrons from the VB will populate states in the CB closer to the $CB_{min}$, reducing the effects of Pauli blocking on XUV absorption. In the same way, the larger density of states induced by BSo will favor the CL transition for $E_{ph}>E_{CL}$ (Fig. 3(f)). As a consequence, in the post-edge case, BSo will produce an increase of sample absorption (measurable as a decrease of the sample transmission). Moreover, the larger number of generated core-holes will increase the probability of XES transitions from both the CB and VB, since a larger number of final states would be available. The above described phenomenology of BSo with: (1) a smaller decrease of sample XAS with respect to the Pauli blocking process in the pre-edge region; and (2) an increase of both sample absorption and spontaneous emission rate in the post-edge region; fit well with the observed trend in the dynamics of unoccupied (XAS) and occupied (XES) states for c-Si $L_{2,3}$ edge reported in Fig. 2(a) and 2(c). Zijlstra et al. identify this BSo process as a precursor of the non-thermal melting of c-Si.

However, Zijlstra's investigation refers to the electronic dynamics occurring in the first picosecond after the pump. We instead explored the evolution of the system on the time scale of thermal phenomena driven by electron-phonon coupling: we observed a relatively fast recovery in the first 25 (50) ps, followed by slower dynamics. The excited and delocalized electrons created by the pump pulse are expected to occupy nonbonding and antibonding states in the conduction band and to weaken the atomic covalent bonds. In this time delay range we expect lattice modifications to take place during the system thermalization. Despite the different pumping fluence, we note that the ten-to-hundreds of picoseconds recovery dynamics we observe is similar to that reported by Beye et al. after the liquid-liquid phase transition.

Also, in the XES case, we observed a ~50 ps recovery time, followed by a slower dynamic. From the latter observation we might infer that transmission and emission dynamics are related to the same electronic/structural phenomenon in the excited c-Si in the first 50 ps after the laser pulse. A detailed description of this specific phenomenon is beyond the scope of this work. Nevertheless, we think that our correlation spectroscopy approach provides great advantages in the investigation of electronic dynamics, due to its capability of extracting information from both occupied and unoccupied electronic states simultaneously.

## IV. CONCLUSION

In this work we presented the implementation of correlation stochastic spectroscopy in the EUV regime, using SASE-like FEL pulses. We demonstrated the potential of this approach to overcome the efficiency issues of the FEL scan-based spectroscopy taking advantage of the spectral instability of SASE-like radiation. This allows us to reconstruct high energy resolution XAS/XES spectra of c-Si. We used this approach to explore the dynamics of c-Si excited by optical radiation in a fluence regime below that needed to induce non-thermal melting. We interpreted the experimental findings in terms of the bond-softening phenomenon, whose interest lies in the dynamic interplay between atomic bonding nature and electronic structure. We revealed that the loss of covalent bonding due to an optical excitation (392 nm wavelength ultrashort pulses with 20 mJ/cm$^2$ fluence) results in a distortion of the c-Si absorption edge, different from the one resulting from conventional heating of the Fermi-Dirac distribution [58, 59], and also influences XES. This observation, while confirming the non-thermal nature of the photo-induced phase transformation in c-Si, indicates a rearrangement of the crystalline bonds occurring on the timescale of electron-phonon coupling.

Because of the inherent dynamic nature of this process, a time-resolved investigation technique is essential, and stochastic spectroscopy has proven to be effectively suited to this task.

In a more general context, EUV stochastic correlation spectroscopy provides comparable information with synchrotron-based measurements, in terms of acquisition time, signal-to-noise ratio and energy resolution, with the valuable additional benefit of allowing time-resolved measurements with sub-ps resolution and the simultaneous determination of XES response.

EUV correlation spectroscopy can be applied to dynamical studies of elastic and inelastic response in soft-matter samples, such as for example the selective enhancement of Raman modes in the aromatic ring of enantiomers [60, 61]. By adding the control over the dichroic response, EUV correlation spectroscopy could also be exploited to study, e.g., the dynamics of spin magnons in antiferromagnetic structures [62], where high energy resolution is necessary to extract valuable information on such inelastic processes.

Finally, it is worth mentioning that correlation-based spectroscopy techniques combined with sub-fs resolution might represent a valuable tool for attosecond science. Indeed, the majority of the Tr-XAS/XES setups use a monochromator, which inherently results in a pulse lengthening, and therefore can deteriorate the time resolution. On the other hand, it has been shown that approaches based on correlating the energy fluctuations within the emission bandwidth of a sub-fs pulse allows the statistical retrieval of the energy resolution, while potentially maintaining the sub-fs time resolution [63].

The authors declare no competing interests.

## ACKNOWLEDGMENTS

We acknowledge FERMI staff for the support during the data collection campaign, in particular Luca Giannessi and Giuseppe Penco for the discussion on the optical klystron lasing scheme.

J.S acknowledges partial support from the National Science Centre (Poland) under grant No. 2017/27/B/ST2/01890.